\documentstyle[11pt,aaspp4]{article}

\begin{document}

\title{Optical Appearance of the Debris of a Star\\
Disrupted by a Massive Black Hole}
\author{Abraham Loeb$^1$ and Andrew Ulmer$^2$}
\medskip
\affil{1. Astronomy Department, Harvard University, 60 Garden St.,
Cambridge, MA 02138}
\medskip 
\affil{2. Princeton University Observatory, Peyton Hall, Princeton, NJ 08544}
  
\begin{abstract}
We show that the disruption of a star by a $\sim 10^6M_\odot$ black hole in
a galactic nucleus could under favorable circumstances produce an
optically--thick envelope that radiates with a thermal spectrum at the
Eddington limit, $\sim 10^{44}~{\rm erg~s^{-1}}$, for tens of years.  The
low apparent temperature of this envelope, $\sim 10^4$K, would be easily
detectable in optical surveys.  If most galaxies harbor a massive black
hole at their center, then the Sloan Digital Sky Survey might find hundreds
of galaxies with nuclear activity of this type.  Because the envelope is
driven to shine near the Eddington limit, a measurement of the source
redshift and total luminosity could yield the black hole mass.

\end{abstract}


\centerline{submitted to {\it The Astrophysical Journal}, March 1997}

\vfil
\eject

\section{Introduction}

At present there is strong evidence, based on stellar kinematics and gas
dynamics, for the existence massive black holes in the centers of most
nearby galaxies (Rees 1997, and references therein).  One of the signatures
of a $\ga 10^6M_\odot$ black hole in a galactic core is that stars on
nearly radial orbits would be tidally disrupted by the black hole every
$\sim 10^4{\rm yr}$.  The impact parameter required for the disruption of
solar--type stars is small, $\sim 25 (M_{\rm bh}/10^6M_\odot)^{-2/3}$
Schwarzschild radii, where $M_{\rm bh}$ is the black hole mass.  As a
result of the disruption process, about half of the stellar mass remains
bound.  After the first passage of the star, the debris forms an elongated
stream of gas which later spreads out, intersects itself, and dissipates
its orbital kinetic energy into heat (Lacy, Townes,
\& Hollenbach 1982; Rees 1988; Evans
\& Kochanek 1989; Canizzo, Lee, \& Goodman 1990; 
Laguna et al. 1993).
Because of the large range of densities and distance scales involved,
the details of this supersonic dissipation process are, by and large,
still uncertain (for recent discussions, see Khokhlov, Novikov,
\& Pethick 1993;  Kochanek 1994; Frohlov et al.  1994).
Generally, one expects the dissipation of the orbital energy to last
several dynamical times ($\sim$ months) and to eventually arrange the gas
in an optically--thick cloud around the black hole.  Since much of the
debris is marginally bound, the dissipation of orbital energy heats the gas
close to its virial energy.  Near pericenter, shocks may organize the high
entropy debris into a thick rotating torus, which would be dominated by
radiation pressure (Rees 1988).  In this process some material is
ejected out of the orbital plane.  Further away from pericenter, rotation
is less important and the strong radiation pressure, could disperse the
marginally bound gas into a quasi--spherical configuration.  The accretion
of gas from the bound debris cloud onto the black hole releases energy in
radiation and defines the observable signature of this event.  For the
purpose of designing observing strategies for tidal disruption events, it
is of fundamental importance to know the expected luminosity and spectral
band of the resulting radiation.

A radiation--pressure dominated torus which accretes onto a massive black
hole, is expected in its simplest form to produce mostly UV photons with
energies $\sim 10^2~{\rm eV}$, and relatively little flux (with a
bolometric correction of $\sim$7 mag) in the optical band (Ulmer 1997).
Based on this fact, one might conclude that {\it optical} detection of
disruption events is a difficult task.  In this paper we show, however,
that much of the radiation emitted by the inner torus could be processed
through a surrounding gaseous envelope, and finally be seen by an external
observer in the optical band.  If the torus shines near the Eddington
limit, then the infall of gas from the surrounding debris cloud would be
moderated by radiation pressure.  The hard UV radiation produced in the
torus would then be processed through a thick layer of gas, degraded to low
photon energies, and eventually emitted from a photosphere in the
optical--UV band.  Since the disrupted star was initially on a nearly
radial orbit, the angular momentum of its debris has little dynamical
significance at large radii, where much of the marginally--bound envelope
resides.  The outer geometry of the envelope would therefore be close to
spherical, whereas its base near the torus would deviate from sphericity
due to rotation (see Fig. 1).  The conversion of debris mass into radiation
in the torus would fuel the surrounding envelope, in much the same way as
nuclear reactions energize stellar interiors. At the same time, the
luminosity of the torus would be controlled by its feeding rate from the
surrounding envelope as long as the envelope contains most of the debris
mass.  Under these circumstances, the surrounding envelope might approach a
steady state, in which the interplay between gravity and radiation pressure
provides a stabilizing feedback; a fluctuation which increases the
luminosity above the Eddington limit would result in an outflow and hence
reduce the accretion luminosity, while a sub--Eddington fluctuation would
increase the infall rate and hence bring the luminosity back to its
equilibrium value (Cowie, Ostriker, \& Stark 1978).  The structure of such
an envelope is simplified by the fact that a fully ionized gas which is
dominated by Thomson opacity and radiation pressure tends to approach a
uniform entropy state. Since gravity is fixed by the massive black hole,
the envelope will then obtain a universal (polytropic) density profile.
The observed properties of this envelope will depend only on the masses of
the black hole and the debris.  Furthermore, the Eddington luminosity
depends solely on the black hole mass, and if an approximate distance were
known to the source, (e.g.  through a redshift), then the black hole mass
could be measured.

The geometry of the above cofiguration is similar to that of a
Thorne-\.{Z}ytkow Object (T\.{Z}O, Thorne \& \.{Z}ytkow 1975), which is a
red supergiant with a neutron star core.  However, there are a number of
important physical differences between the systems.  In the envelope we
consider, the gravitational force is fixed by the central point mass
whereas the envelope of a T\.{Z}O provides most of its total mass.
Secondly, the interior temperatures and densities in our models are not
sufficiently high to produce pairs or significant nuclear fusion, as is the
case for T\.{Z}Os with envelopes more massive than $\sim 10 M_\odot$. Our
envelopes are powered by gravitational energy, more similarly to low mass
T\.{Z}Os.

The existence of a steady, spherical, optically--thick envelope around
the black hole should be regarded as the most optimistic expectation for
the optical appearance of a disruption event.  Our model can therefore be
used to motivate searches for the most luminous disruption events in the
local population of galaxies.

The outline of this paper is as follows.  In \S 2 we derive the universal
density profile and effective temperature of an Eddington envelope around
a massive black hole.  We examine the self--consistency condition for
the existence of this steady state and estimate its lifetime. We show that
the required radiative efficiency can naturally be supplied by an accreting
torus near the black hole.  Section 3 examines the observational signatures
of the Eddington envelopes.  Finally, \S 4 summarizes the main
conclusions of this work.

\section{Structure of an Eddington Envelope around a Massive Black Hole}

We consider the envelope of high--entropy gas that might form around a
massive black hole of mass $M_{\rm bh}$ as a result of the tidal disruption
of a star.  The envelope is dominated by Thomson opacity, because at its
characteristically low densities ($\lesssim 10^{-12}$ g~cm$^{-3}$), the
bound--bound, bound--free, and free--free opacities are relatively
unimportant (Lamers \& Burger 1989). We denote the mass of the stellar
debris that is bound to the black hole by $M_\star$ ($\sim$ half of the
initial mass of the star).  Our underlying assumption is that the cooling
time of the bound gas is much longer than its dynamical time, so that the
envelope relaxes to a steady--state configuration. The self--consistency of
this assumption will be demonstrated later.  We also assume that radiation
pressure dominates over gas pressure, because of the high temperature
achieved through dissipation of the orbital energy of the marginally bound
debris.  Based on the equations of hydrostatic equilibrium and radiative
transfer, steady radiative envelopes of this type must have a constant
ratio of gas pressure to total pressure or a uniform entropy (Eddington
models).  With radiation domination and uniform entropy, the radiation
pressure, $p_{\rm rad}$, is expressed in terms of the mass density of the
gas $\rho_{\rm gas}$,
\begin{equation}
P_{\rm rad}= K \rho_{\rm gas}^{4/3},
\label{eq:1}
\end{equation}
where $K=$const is related to the entropy of the gas.
The hydrostatic equilibrium equation
\begin{equation}
{GM_{\rm bh}\over r^2}=-{1\over \rho_{\rm gas}}{\partial P_{\rm rad}
\over \partial r}=-4K{\partial \rho_{\rm gas}^{1/3}\over
\partial r},
\label{eq:2}
\end{equation}
admits the solution,
\begin{equation}
\rho_{\rm gas}=\left({GM_{\rm bh}\over 4K}\right)^3 {1\over r^3}.
\label{eq:3}
\end{equation}
The total mass of the envelope is then,
\begin{equation}
M_\star=4\pi \int_{R_{\rm in}}^{R_{\rm out}} \rho_{\rm gas} r^2 dr=
\left({GM_{\rm bh}\over 4K}\right)^3 4\pi \ln\left({R_{\rm out}
\over R_{\rm in}}\right),
\label{eq:4}
\end{equation}
and so we may write
\begin{equation}
\rho_{\rm gas}= \left[{M_\star\over 4\pi\ln({R_{\rm out}
/R_{\rm in}})}\right]{1\over r^3},
\label{eq:5}
\end{equation}
where $R_{\rm out}$ and $R_{\rm in}$ are the inner and outer radii
of the envelope.
We set the inner radius to be of order the
pericenteric radius for the parent star orbit. The condition 
for tidal disruption implies,
\begin{equation}
R_{\rm in}\approx R_\star \left({M_{\rm bh}\over M_\star}\right)^{1/3}
\sim 10^{13} \left({M_{\rm bh}\over 10^6M_\odot}\right)^{1/3}~{\rm cm}~,
\label{eq:9}
\end{equation}
where $R_\star\sim R_\odot$ is the initial radius of the disrupted star.
The exact value of $R_{\rm in}$ is of little importance since it enters
only logarithmically into our final results.

The hydrostatic equilibrium equation~(\ref{eq:1})
holds only out to the radius where the radiation is trapped,
and so we associate the outer radius of the envelope with 
the photospheric condition, 
\begin{equation}
{\sigma_{\rm T}\over \mu_e m_p}\int_{R_{\rm out}}^\infty \rho_{\rm gas} dr =1,
\label{eq:6}
\end{equation}
where $\sigma_{\rm T}$ is the Thomson cross--section, 
$m_p$ is the proton mass, and $\mu_e$ is the mean atomic
weight per electron.
For a hydrogen mass fraction, $X=0.74$,
$\mu_e\approx 2/(1+X)=1.15$.
This implies,
\begin{equation}
R_{\rm out} \approx \left[ {\sigma_{\rm T} M_\star\over
8\pi \mu_e m_p \ln(R_{\rm out}/R_{\rm in})}\right]^{1/2}\approx
1.7\times 10^{15}\left({M_\star\over 0.5 M_\odot}\right)^{1/2}~{\rm cm}.
\label{eq:7}
\end{equation}
This result is only approximate, since equation~(\ref{eq:5}) does not
strictly hold past the photosphere. Nevertheless, the steep dependence of
the optical depth on radius ($\propto r^{-2}$) interior to the photosphere,
implies that the actual location of the outer radius could not differ much
from this value.

The flux of radiation at each radius is obtained from the radiative transfer
equation
\begin{equation}
F(r)\equiv{L\over 4\pi r^2}=-{\mu_e m_p c\over \sigma_{\rm T} \rho_{\rm gas}}
{\partial P_{\rm rad}\over \partial r},
\label{eq:8}
\end{equation}
and so equation~(\ref{eq:2}) implies that the luminosity of the envelope
equals the Eddington limit,
\begin{equation}
L=L_{\rm E}\equiv {4\pi G \mu_e m_p c M_{\rm bh}\over \sigma_{\rm T}}
=1.4\times 10^{44} \left({M_{\rm bh} \over 10^6M_\odot}\right)~
{\rm erg~s^{-1}}.
\label{eq:19}
\end{equation}

Equation~(\ref{eq:2}) ignored the gas pressure, $P_{\rm gas}=\rho_{\rm gas}
k_{\rm B}T/(\mu_pm_p)$, relative to the radiation pressure,
$P_{\rm rad}= {1\over 3}a T^4$;
to make it exact, its left hand side should be multiplied by
$(1-\beta)$, where
\begin{equation}
\beta\equiv {P_{\rm gas}\over P_{\rm gas}+P_{\rm rad}} \approx 
{k_{\rm B}\over \mu_p m_p}
\left[{48M_\star \over \pi a (GM_{\rm bh})^3 
\ln(R_{\rm out}/R_{\rm in})}\right]^{1/4}
= 10^{-4}\left({M_{\rm bh}\over
10^6M_\odot}\right)^{-3/4} \left({M_\star\over 0.5M_\odot}\right)^{1/4},
\label{eq:beta}
\end{equation}
and $\mu_p\equiv [1/\mu_e+(1+3X)/4]^{-1}= 0.6$. As a result, the luminosity
deviates slightly from the Eddington limit, $L=L_{\rm E}(1-\beta)$.

The luminosity can also be expressed in terms of the effective temperature
of the photosphere $T_{\rm eff}$,
\begin{equation}
L_{\rm E}=4\pi R_{\rm out}^2 \sigma T_{\rm eff}^4,
\label{eq:10}
\end{equation}
yielding,
\begin{equation}
T_{\rm eff}\approx 1.3\times 10^4~{\rm K} \left({M_{\rm bh}\over
10^6 M_\star}\right)^{1/4}.
\label{eq:teff}
\end{equation}
The effective temperature has a very weak dependence on the mass ratio
between the black hole and its envelope and is associated
with the optical--UV band.
The theoretical prediction of the color of
such an envelope is therefore robust.
The interior temperature profile
of the envelope scales as $T\propto \rho_{\rm gas}^{1/3}
\propto r^{-1}$ and reaches a value $\sim 10^6~{\rm K}$
at $\sim R_{\rm in}$.  Note that the internal energy of the envelope,
\begin{equation}
E_{\rm int}\equiv 3\int_{R_{\rm in}}^{R_{\rm out}} P_{\rm rad}~
4\pi r^2 dr = {3\over 4\ln(R_{\rm out}/R_{\rm in})}
{G M_{\rm bh} M_\star\over  R_{\rm in}},
\end{equation}
could naturally be provided by the disruption of a star at the tidal
radius, $\sim R_{\rm in}$.

So far we have ignored the angular momentum of the envelope.
Because the
disrupted star must be on a nearly radial orbit, rotation is unimportant at
large radii, $\sim R_{\rm out}$.  However the centrifugal force could
support the gas against gravity near $R_{\rm in}$.
The  accretion flow onto the black hole converts a fraction
$\epsilon$ of the accreting mass into radiation.
This radiation would then
be reprocessed through the surrounding optically--thick envelope before
it reaches the observer.
Energy balance implies,
\begin{equation}
L_{\rm E}=-\epsilon {\dot M}_\star c^2.
\label{eq:11}
\end{equation}

The radiative efficiency, $\epsilon$, depends on the accretion configuration
around the black hole, where most of the gravitational binding energy of
the gas is dissipated. The envelope always shines at the Eddington limit,
irrespective of the value of $M_\star$, and so ${\dot
M_{\star}}=$const for a steady radiative efficiency, $\epsilon=$const. We
therefore obtain,
\begin{equation}
M_\star(t)=M_\star(0)\left(1-{t\over t_{\rm E}}\right) ~~~~~~~~
{\rm for~0\leq t\leq t_{E}},
\label{eq:12}
\end{equation}
where
\begin{equation}
t_{\rm E}= 20~{\rm yr} \left({M_\star(0)\over 0.5 M_\odot}\right)
\left({M_{\rm bh}\over 10^6M_\odot}\right)^{-1}
\left({\epsilon\over 10\%}\right).
\label{eq:13}
\end{equation}
This time dependence determines the evolution of the effective temperature
in equation~(\ref{eq:teff}), $T_{\rm eff}\propto (1- t/t_{\rm E})^{-1/4}$.

The above envelope configuration could exist only as long as the evolution
time $t_{\rm E}$ is much longer than the system's dynamical time over which
it relaxes to hydrostatic equilibrium,
\begin{equation}
t_{\rm dyn}=\left({GM_{\rm bh}\over R_{\rm out}^3}\right)^{-1/2}
\approx 0.2~{\rm yr} 
\left[{M_{\star}\over 0.5M_\odot}\right]^{3/4}
\left({M_{\rm bh}\over 10^6M_\odot}\right)^{-1/2}.
\label{eq:14}
\end{equation}
Equations~(\ref{eq:13}) and~(\ref{eq:14}) imply that our discussion is
self--consistent as long as the radiative efficiency is sufficiently high,
\begin{equation}
\epsilon\ga 10^{-3} 
\left[{M_\star(0)\over 0.5 M_\odot}\right]^{-1/4}
\left({M_{\rm bh}\over 10^6M_\odot}\right)^{1/2}.
\label{eq:15}
\end{equation}
The existence of an Eddington envelope relies on the fact that the
intersection and dissipation of the debris streams would yield a
super--Eddington mass accretion rate.  This is likely to be the case if
$t_{\rm dyn}\ll t_{\rm E}$ and the black hole mass is low $M_{\rm bh} \la
10^7M_\odot$ (Ulmer 1997).

The radiative efficiency could obtain high values only if the viscous time,
$t_{\rm vis}$, for the transport of angular momentum near the base of the
envelope (where most of the radiation is produced), is comparable or longer
than the photon diffusion time out of this region, $t_{\rm rad}$ (Begelman
\& Meier 1982).  If, however, $t_{\rm vis}\ll t_{\rm rad}$ then most of the
radiation will be trapped and carried with the gas into the black hole,
hence leading to a low $\epsilon$.  Under such circumstances, the disk
luminosity $L\ll L_{\rm E}$ would be unable to support the surrounding
envelope against gravity, and the distant gas will be drained down to
$R_{\rm in}$, where angular momentum becomes important, on its free--fall
time.  To estimate the central viscous time we assume that the base of the
envelope at $r=R_{\rm in}$ is rotationally supported.  By parameterizing
the viscosity coefficient through the relation, $\eta\equiv
\alpha P_{\rm rad}~(r^3/GM_{\rm bh})^{1/2}$ (Shakura \& Sunyaev
1973), we get
\begin{equation} 
t_{\rm vis} \approx {\rho_{\rm gas} R_{\rm in}^2\over
\eta} \sim {4\over \alpha} 
\left({R_{\rm in} \over GM_{\rm bh}/c^2}\right)^{1/2} {R_{\rm in}\over c}
\label{eq:16}
\end{equation}
where we have used equations~(\ref{eq:2}) and~(\ref{eq:5}) to substitute
$P_{\rm rad}/\rho_{\rm gas}\approx GM_{\rm bh}/4R_{\rm in}$ at the base of
the envelope.
For $M_\star=0.5 M_\odot$ and $M_{\rm bh}=10^6M_\odot$,
this yields, $t_{\rm vis}\approx 0.3 (\alpha/10^{-3})^{-1}~{\rm yr}$.
On the other hand, the photon diffusion time is of order, $t_{\rm rad}\sim
\tau R_{\rm in}/c$, where $\tau$ is the optical depth to electron
scattering.  Equations~(\ref{eq:5}) and~(\ref{eq:7}) yield $\tau\approx
0.5(R_{\rm out}/R_{\rm in})^2\sim 10^4$, and $t_{\rm rad}\sim 0.1~{\rm
yr}$.  Thus, for $\alpha\la 3\times 10^{-3}$, $t_{\rm rad}\la t_{\rm vis}$,
and the gas would develop a rotationally supported configuration near
$R_{\rm in}$ with a high radiative efficiency.  However, for $\alpha\gg
3\times 10^{-3}$ most of the radiation would be advected into the black
hole (Narayan 1996) and the stellar debris would disappear within several
months; the advection would be most effective in a spherical geometry, but
less so in a disk geometry for which $\epsilon$ is reduced only by a factor
$\sim t_{\rm vis}/t_{\rm rad}$. Finally, we note that any variability on
short timescales due to thermal or dynamical instabilities near the center
(Chen et al.  1995), would be moderated by the long diffusion time of the
photons through the surrounding optically--thick envelope.

The viscosity parameter is also limited from below so as to yield $t_{\rm
vis}\la t_{\rm E}$. This requires $\alpha\ga 10^{-5} (M_{\rm
bh}/10^6M_\star)(\epsilon/10\%)^{-1}$, based on equations~(\ref{eq:13})
and~(\ref{eq:16}).

For thick accretion tori with radial hydrostatic equilibrium (e.g.
Paczy\'nski \& Wiita 1980 (PW); Jaroszy\'nsky, Abramowicz, \& Paczy\'nski
1980), there is no significant advection of thermal energy.  Because the
inner edge of the torus is pushed by pressure inwards of the marginally
stable Schwarzschild orbit at 6$GM_{\rm bh}/c^2$ and brought to a binding
energy closer to zero, the radiative efficiency of a thick disk is lower
than that of a thin disk.  Using a thick disk code developed by Ulmer
(1997), we investigate the minimum radiative efficiencies for a wide range
of thick disks.  Our disks are embedded in a pseudo--Newtonian potential,
which is a convenient substitute for the general relativistic metric of a
Schwarzschild black hole (PW).  We assume a uniform entropy equal to that
of the Eddington envelope, with pressure described by the polytrope
relation~(\ref{eq:1}).  We further require that the torus join the envelope
smoothly so as to have radial force balance at the transition
radius.\footnote{We have not attempted the much more difficult problem of
requiring force balance along the axial direction in the transition region.
This problem involves the interaction between the radiation emitted by one
wall and the surface of the opposite wall along the funnel of the torus, as
well as the two--dimensional transfer of the emergent radiation through the
ambient envelope.} Under these assumptions, a radius of the last stable
orbit (which lies inside of 6$GM_{\rm bh}/c^2$ due to pressure gradients)
and a run of angular momentum, uniquely define the disk.

In Figure 2, we show the results of a search through parameter space for the
minimum efficiency disks, namely those with the inner radius closest the
marginally bound orbit.
We have used angular momentum distributions of the form $j(r) = A(b) r^{b}$
in Model 1, and
\begin{equation}
j(r)  = 
\left\{\matrix{
 A(b) r^{b} j_{_{\rm Kep}}(r)
\hfill&({\rm for}~r>r_{\rm b})\hfill\cr
j(r_{\rm b}) 
\hfill&({\rm for}~r < r_{\rm b})\hfill,\cr}\right. 
\end{equation}
in Model 2. Here $j_{_{\rm Kep}}$ is the Keplerian angular momentum per
unit mass in the PW potential, and the constant, $r_{\rm b} = 1/(1+b)$,
ensures that $dj/dr \ge 0 $ for all $r$.  All radial variables are in units
of the Schwarzschild radius.  Because the PW potential is an approximation,
the efficiencies can be slightly higher (6.3~\%) than the maximum
efficiency for a thin disk in the Schwarzschild metric (5.7~\%).  We
require that the specific angular momentum of the torus be as large as the
specific angular momentum of the star before its disruption at the tidal
radius.

The radiative efficiencies found for this broad class of models are all
significantly higher than the value of $10^{-3}$, required for the
existence of an envelope [Eq.~(\ref{eq:15})]. The efficiencies are expected
to be even higher if the black hole is spinning (Thorne 1974), a generic
situation for a black hole that grew through accretion from a disk (Bardeen
1970).

For super--Eddington thick disks, the luminosity scales logarithmically
with mass accretion rate (Paczy\'nski 1980), whereas for sub--Eddington
luminosities, the luminosity scales linearly with mass accretion rate.
Consequently, the luminosity will never be highly super--Eddington.  If the
disk luminosity falls below the Eddington limit, the mass supply rate to
the disk will increase, giving rise to a stabilizing feedback near the
Eddington limit.  Such a feedback is likely to be most effective when the
disk mass is small relative to the envelope and the mass infall from the
envelope controls the disk accretion rate.  However, it is also possible
that the disk will not respond quickly enough to the envelope, and that the
envelope would be pushed--out impulsively by a super--Eddington eruption in
the disk.  If sufficient energy is stored in the disk, then such an event
could lift the envelope and unbind it.  Even under steady state conditions,
the outer boundary of the envelope will inevitably develop a
radiation--driven wind, in analogy with hot stars (Kudritzki et al. 1989,
and references therein).  In this case, the wind would be accelerated by
the extra radiative force at the photosphere due to atomic lines.

Irrespective of the details of the radiative force, the mass loss rate in
the wind, ${\dot M}_{\rm w}$, is limited by momentum conservation,
\begin{equation}
{\dot M}_{\rm w} v_{\infty}= f_{\rm w} {L/ c},
\label{eq:22}
\end{equation}
where $f_{\rm w}\la 1$ (Cassinelli 1979; but see the effects of multiple
scatterings in Lucy \& Abbott 1993), $v_{\infty}$ is the terminal
wind velocity, and $L/c$ is the total momentum output per unit time carried
by the radiation.  In hot stars with luminosities near the Eddington limit,
this relation is obeyed with $f_{\rm w}\sim 1$.  Based on the previous
paragraph, we distinguish between two cases: (i) slow wind out of a
steady--state envelope [$L=L_{\rm E}(1-\beta)$]; and (ii) fast wind in a
transient envelope [$L\ga L_{\rm E}$].  In both cases we scale
$v_{\infty}\equiv
\omega_{\rm w} (2GM_{\rm bh}/R_{\rm out})^{1/2}$.  In the slow wind case,
the wind velocity is of order the net escape speed from the photosphere
(cf.  Figure 15c in Kudritzki et al.  1989), i.e.  $\omega_{\rm w}\approx
\beta^{1/2}\approx 10^{-2}$ [cf. Eq.~(\ref{eq:beta})], leading to
$v_{\infty}\approx 40~{\rm km~s^{-1}}$.  Since the wind velocity is small,
the evaporation time of the envelope is limited in this case by its long
crossing time,
\begin{equation}
t_{\rm w}\ga {R_{\rm out}\over v_{\infty}}\approx \beta^{-1/2} t_{\rm dyn}
= 20~{\rm yr}\left({M_\star\over 0.5M_\odot}\right)^{5/8}
\left({M_{\rm bh}\over 10^6M_\odot}\right)^{-1/8}.
\label{eq:slowwind}
\end{equation}
In the fast wind case, where $(L/L_{E}-1)$ is of order unity for a period
lasting more than $\sim R_{\rm out}/v_\infty$ (e.g. due to an eruption in
the disk), $\omega_{\rm w}\sim 1$ so that $v_{\infty}\approx 4\times
10^3~{\rm km~s^{-1}}$, and the lifetime of the envelope is
\begin{equation}
t_{\rm w}= {M_\star \over {\dot M}_{\rm w}}\approx 5~{\rm yr}
\left({M_\star\over 0.5 M_\odot}\right)^{3/4}
\left({M_{\rm bh}\over 10^6M_\odot}\right)^{-1/2}
\left({f_{\rm w}\over 0.5}\right)^{-1}.
\label{eq:23}
\end{equation}
Since the photon diffusion time $t_{\rm rad}\sim 0.1~{\rm yr}$ is much
shorter than $t_{\rm w}$, a transient super--Eddington flux which lasts for
several months ($\sim R_{\rm out}/v_{\infty}$) due to unsteady accretion in
the disk, will only ablate the outermost layer of the envelope.  Based on
equations~(\ref{eq:13}), (\ref{eq:slowwind}), and~(\ref{eq:23}), we infer
that the characteristic source lifetime is between several years and
several decades.

\section{Observational Signatures}

An Eddington envelope of a $\sim 10^6M_\odot$ black hole would shine with
an optical luminosity $\sim 4\times 10^{10}L_\odot$, comparable to the
total luminosity of a bright galaxy. It should therefore be straightforward
to detect an unresolved nuclear emission of this type in any galaxy.

The emission events are transients with typical durations of years to
decades.  Since the disruption rate of stars is $\sim 10^{-4}~{\rm
yr}^{-1}$ per galaxy (Rees 1988), a fraction $\sim 10^{-4}$--$10^{-3}$ of
all galaxies should show signatures of a disruption event at any given
time. This is $\sim 1$--$10\%$ of all Seyfert galaxies (Huchra \& Burg
1992), which are active galactic nuclei with comparable luminosities.  Most
Seyferts show a UV excess consistent with thermal emission at a temperature
$\ga 10^5$K (Elvis et al. 1994), well above our predicted photospheric
temperature [Eq.~(\ref{eq:teff})]; this agrees with our expectation that
Eddington envelopes should reflect a small minority of all Seyferts.  The
forthcoming Sloan Digital Sky Survey (Gunn \& Knapp 1993; see also
http://www.astro.princeton.edu/BBOOK) will obtain low--resolution spectra
of $\sim 10^6$ galaxies (and image many more galaxies in five bands), and
might therefore find more than several hundred galaxies with nuclear
activity due to a tidal disruption event. The nuclear emission from these
galaxies would disappear on a timescale of years to decades, and could
therefore be identified through repeated observations, analogous to those
of supernova searches (e.g. Perlmutter et al. 1996). Follow--up
observations with the Hubble Space Telescope could be used to extend the
measured spectra of these objects into the UV, and to confirm their low
emission temperatures.

Since the Eddington envelopes are very optically thick ($\tau \ga 10^4$),
their emission spectrum is to first order thermal.  The details of this
spectrum resemble those of the most luminous, extended blue supergiants
(of which Luminous Blue Variables are a subclass)
which have luminosities approaching the Eddington limit (Humphreys \&
Davidson 1994).
In Figure 3 we show results from the stellar atmosphere
code ATLAS12 (Kurucz 1995) for the spectra of a nearly Eddington
envelope with $T_{\rm eff}= 1.3\times 10^4$~K [cf.  Eq.~(\ref{eq:teff})].
At a fixed effective temperature, the lower the surface gravity, g, is
(i.e. the larger the radius is), the more luminous the envelope becomes.
The log(g)=1.5 (cgs units) curve is the lowest gravity model calculable
with ATLAS12,
and is very close to the Eddington limit, which occurs at log(g)$\approx
1.3$.
At the above value of $T_{\rm eff}$, low gravity increases the ionization
fraction of hydrogen and weakens the Balmer decrement and absorption lines
relative to supergiant spectra.
For Eddington envelopes, g$=GM_{\rm bh}/R_{\rm out}^2\propto M_{\rm
bh}/M_\star\propto T_{\rm eff}^4$, and so the envelopes of more massive
black holes would have higher log(g) and higher effective temperature.

Although the broad band spectrum of an Eddington envelope resembles that
of a hot star, the lower gravity of the envelope increases the ionization
fraction, and hence changes the fine details of the spectrum. In addition,
the lower densities of the Eddington envelope suppress bound--bound
transition lines which result from atomic collisions. In reality, the
spectrum shown in Figure 3 might be supplemented by broad absorption lines
from the fast wind (with a characteristic velocity width of up to $v_{\rm
w}\sim 10^4~{\rm km~s^{-1}}$) and broad emission lines from the broad line
region of the active nucleus.

The extragalactic nature of the source can be established from the
cosmological redshift of its absorption and emission features.  Due to the
gravitational potential of the black hole, these features should also be
redshifted relative to the narrow emission lines of the host galaxy.  The
corresponding velocity shift,
\begin{equation}
\Delta v= {G M_{\rm bh}\over R_{\rm out} c} = 26~{\rm km~{s^{-1}}}
\left({M_{\rm bh}\over 10^6M_\odot}\right)
\left({M_\star\over 0.5M_\odot}\right)^{-1/2} ,
\label{eq:24}
\end{equation}
is small but possibly detectable. This shift is somewhat larger than the
spectral broadening associated with the thermal velocity of the gas at the
photosphere ($\sim 10~{\rm km~s^{-1}}$).

\section{Conclusions}

In this work we have shown that much of the energy radiated by the debris
of a star disrupted by a massive black hole could be channeled into the
optical--UV band. The photosphere of the debris cloud would then shine at
the Eddington luminosity with an effective temperature of $\sim 10^4
(M_{\rm bh}/10^6 M_\star)^{1/4}$K [cf.  Eq.~(\ref{eq:teff})].

In this case, the disruption event would appear as unresolved emission from
the nucleus of the host galaxy, marked by a thermal spectrum similar to
that of a hot star but with fewer absorption lines (Fig. 3).  If broad
emission lines accompany this nuclear activity, their redshift can serve as
the definitive proof for the extragalactic origin of the associated thermal
emission.  Studies of reverberation mapping imply that the broad line
region is located at a radius $\sim 3\times 10^{16}(L/10^{44}~{\rm
erg~s^{-1}})^{1/2}~{\rm cm}$ (Peterson 1993; Maoz 1996), much larger than
the expected photospheric radius of the debris cloud, $\sim 10^{15}~{\rm
cm}$ [cf. Eq.~(\ref{eq:7})]. However, the existence of broad emission lines
is in question, in particular because any preexisting broad line clouds
might be swept away by the wind of unbound debris from the disruption
event.

Because a long--lived envelope must radiate near the Eddington limit, a
measurement of the source redshift and total flux could yield the black
hole mass [Eq.~(\ref{eq:19})].  The effective temperature of the envelope
could then be used to fix the mass of the envelope [Eq.~(\ref{eq:teff})].
Due to gravitational redshift, the spectral features of the envelope would
be redshifted relative to the host galaxy by $\sim 30~{\rm km~s^{-1}}$
[Eq.~(\ref{eq:24})].

We have shown that the universal solution for the steady envelope is
self--consistent as long as the radiative efficiency near the black hole is
higher than a fraction of a percent.  This condition could naturally be
satisfied by an inner accretion torus (Fig. 2). 

The characteristic lifetime of the above envelope is between years and
decades, irrespective of whether it is determined by the cooling time of
the envelope [Eq.~(\ref{eq:13})] or by the momentum flux limit on a
radiation--driven wind [Eqs. \ref{eq:22})--(\ref{eq:23})].  When combined
with the event rate of $\sim 10^{-4}~{\rm yr^{-1}}$ per galaxy, this
implies that a fraction of $\sim 10^{-4}$--$10^{-3}$ of all galaxies might
show signs of activity due to the disruption of a star in their nucleus.
If our predicted optical appearance of the disruption events is generic,
then the forthcoming Sloan Digital Sky Survey might find several hundreds
of galaxies which show Seyfert--like luminosity in their nucleus ($\sim
10^{44}~{\rm erg~s^{-1}}$) and have thermal spectra with an effective
temperature $\sim 10^4$ K.  These sources would exhibit considerable
dimming over a decade of repeated observations.

\acknowledgements

We thank E. Fitzpatrick, J. Goodman, R. Kurucz, R. Narayan, B. Paczy\'nski,
G. Rybicki, and D.  Sasselov for useful discussions. AL was supported in
part by the NASA ATP grant NAG5-3085 and the Harvard Milton fund. AU was
supported by an NSF graduate fellowship and NSF grants AST93-13620 and
AST95-30478.

\begin{figure}
\plotone{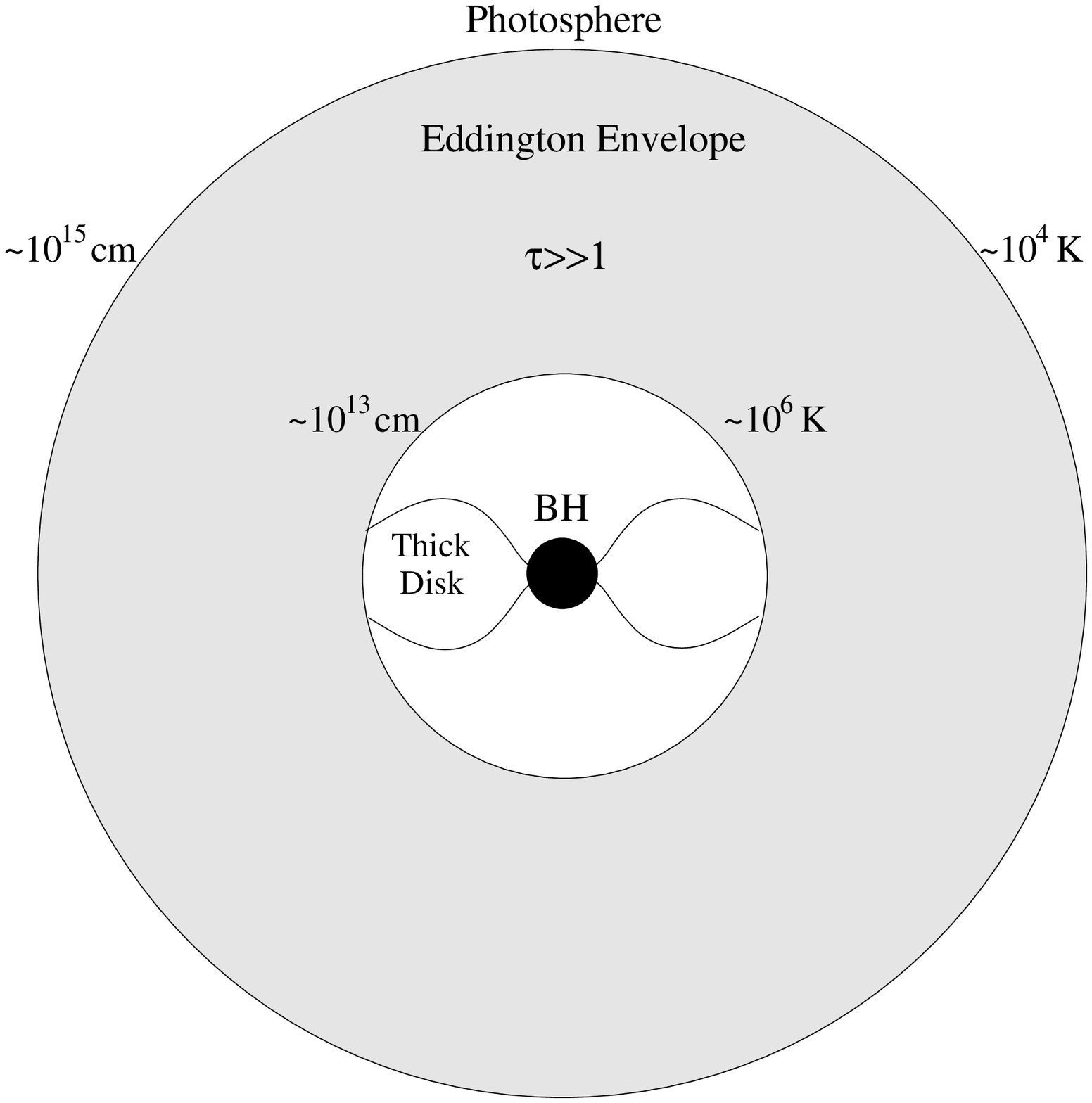}
\caption{A sketch of the possible long--term geometry of the debris
of a disrupted star around a massive black hole. The energy output from the
inner torus is controlled by the mass feeding rate from the surrounding
envelope. If the luminosity exceeds (declines below) the Eddington value
the feeding is reduced (increased) and the luminosity returns to its 
equilibrium value.}
\end{figure}

\begin{figure}
\plotone{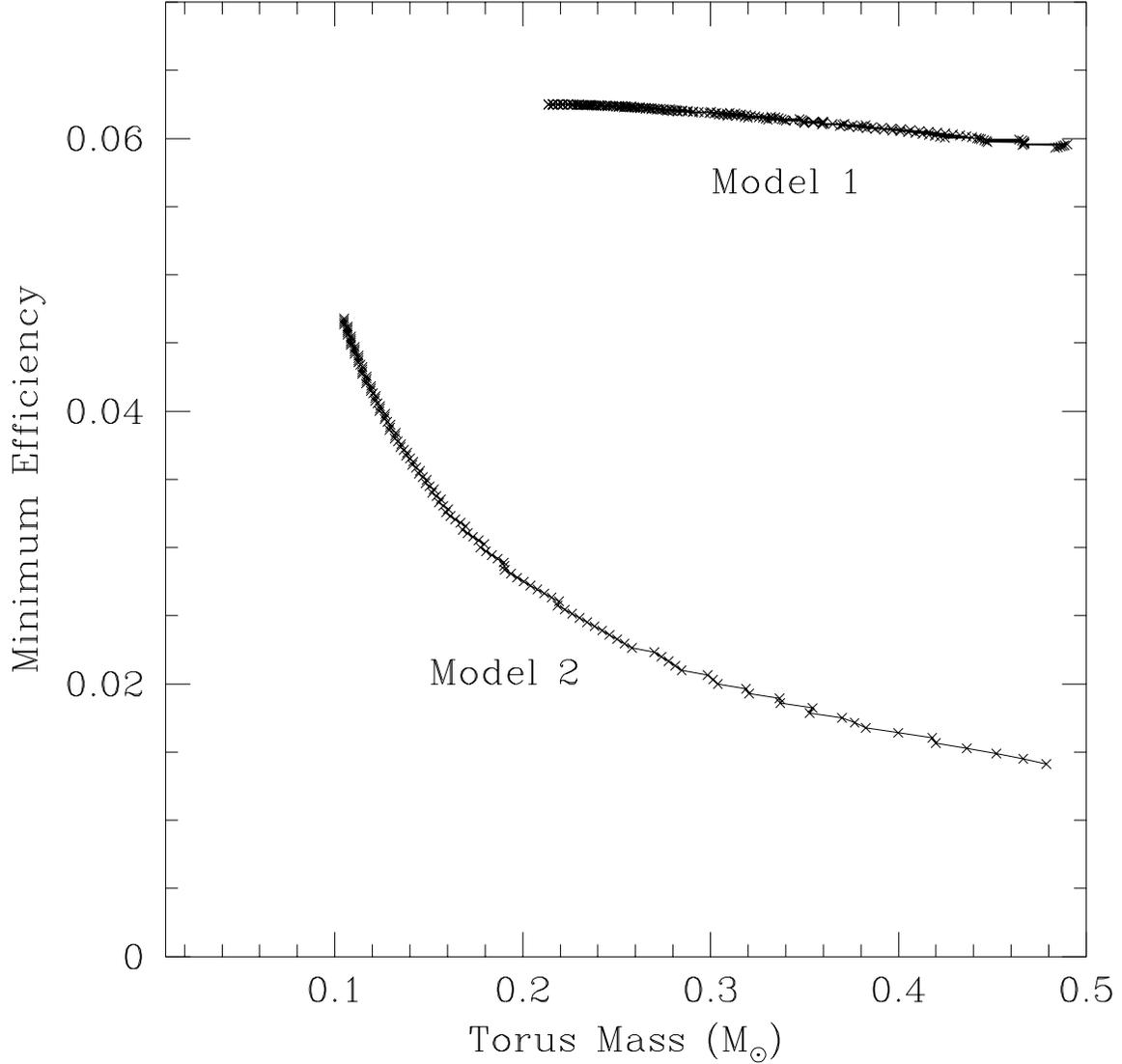}
\caption{Minimum radiative efficiencies for steady--state accretion
in a thick disk located at the tidal radius of the disrupted star.  The
results apply to the disruption of a $1 M_\odot$ star by a 10$^6 M_\odot$
black hole.  The two models involve different functional forms for the
angular momentum distribution, as described around equation (21).}
\end{figure}

\begin{figure}
\plotone{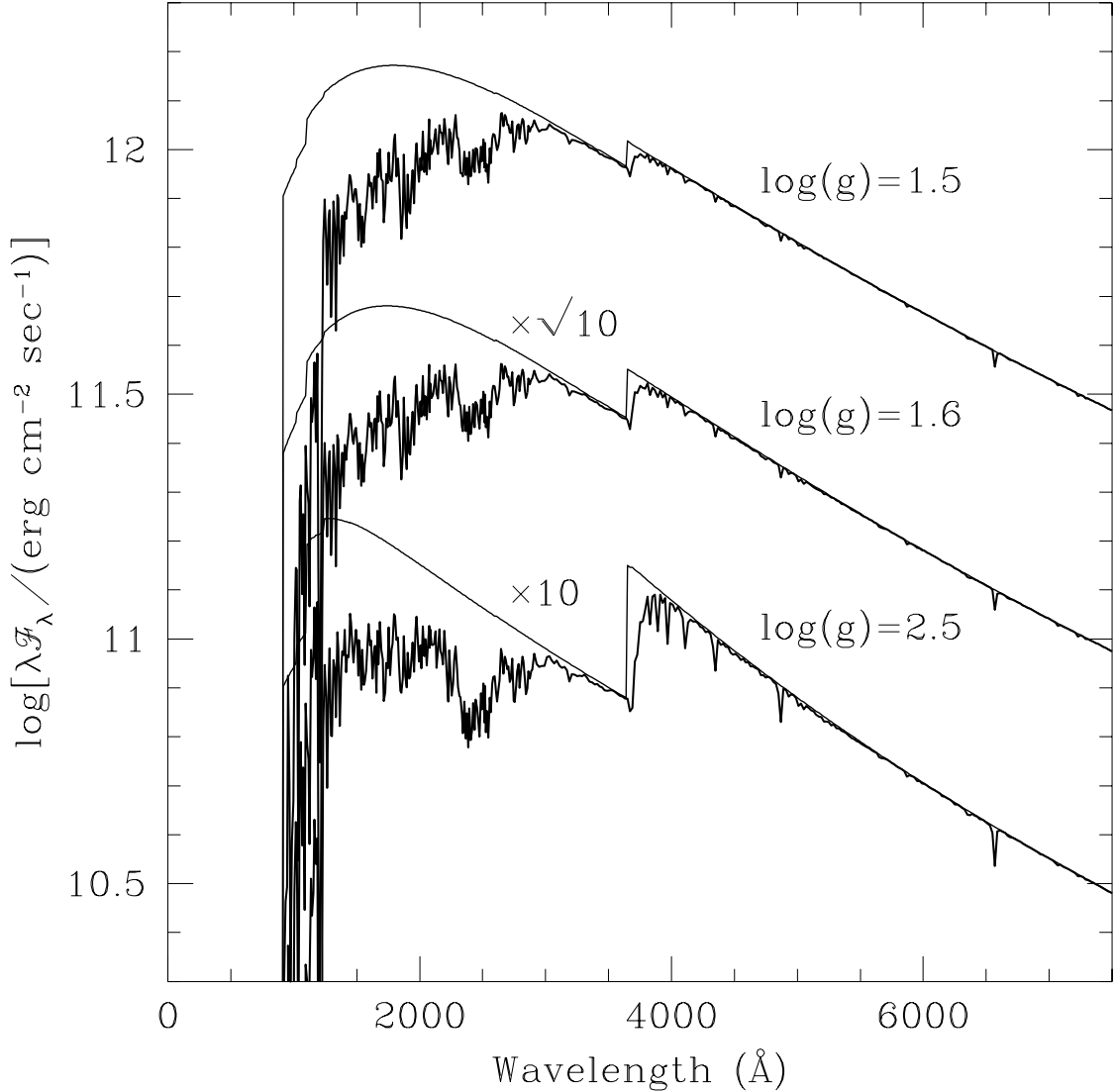}
\caption{Spectral flux times wavelength
${\cal \lambda F_\lambda}~{\rm (ergs~cm^{-2}~s^{-1})}$ at the photosphere,
for model atmospheres calculated with ATLAS12 (Kurucz 1995).  The smooth
lines are the spectra with only continuum hydrogen absorption.  {\bf Top
curve [$\log (\rm{g}) = 1.5$]:} detailed spectrum of a near Eddington
envelope for $T_{\rm eff}= 13,000$~K (which corresponds to ${M_{\rm
bh}=10^6 M_\star}$).  This spectrum is the highest luminosity calculable
with ATLAS12 before the surface points become radiatively unstable [this
limit is the opacity--modified Eddington limit of Lamers and Fitzpatrick
(1988)].  {\bf Middle curve [$\log (\rm{g}) = 1.6$]:} spectral flux
(reduced by a factor of $10^{1/2}$) 
corresponding to the most luminous observed
supergiants which occur at the Humphreys--Davidson limit (Humphreys \&
Davidson 1979).  {\bf Bottom curve [$\log (\rm{g}) = 2.5$]:} spectrum
(shifted by a factor of 10) of a typical supergiant of luminosity class II
or Ib. Note the pronounced Balmer decrement.  All calculations were
performed for a solar metalicity envelope with a large micro--turbulence of
$8~{\rm km~s^{-1}}$.  }
\end{figure}
\end{document}